\documentclass[letterpaper, 10 pt, conference]{ieeeconf}
\IEEEoverridecommandlockouts                            
\usepackage{mathtools,amsmath,amssymb}
\usepackage{color,soul,graphicx,etoolbox}
\usepackage{algorithm2e}

\newcommand{\dd}{\text{d}}
\newcommand{\qiy}{q}

\AtBeginEnvironment{bmatrix}{\setlength{\arraycolsep}{2pt}}

\begin{document}
\title{\huge\bf Long hauling eco--driving: heavy--duty trucks operational modes control with integrated road slope preview \thanks{This research was funded by the European Union’s Horizon 2020 Research and Innovation Programme, Grant Agreement no. 874972, Project LONGRUN.}}

\author{Gustavo R. Gon\c{c}alves da Silva$^{1}$ and Mircea Lazar$^{1}$%
\thanks{$^{1}$Department of Electrical Engineering, Eindhoven University of Technology, The Netherlands: 
{\tt\small g.goncalves.da.silva@tue.nl, m.lazar@tue.nl}}
}

%\author{\IEEEauthorblockN{Gustavo R. Gon\c{c}alves da Silva}
%\IEEEauthorblockA{\textit{Department of Electrical Engineering} \\
%\textit{Eindhoven University of Technology}\\
%Eindhoven, The Netherlands\\
%g.goncalves.da.silva@tue.nl}
%\and
%\IEEEauthorblockN{Mircea Lazar}
%\IEEEauthorblockA{\textit{Department of Electrical Engineering} \\
%\textit{Eindhoven University of Technology}\\
%Eindhoven, The Netherlands\\
%m.lazar@tue.nl}
%}

\maketitle

\begin{abstract}%
In this paper, a complete eco--driving strategy for heavy--duty trucks (HDT) based on a finite number of driving modes with corresponding gear shifting is developed to cope with different route events and with road slope data. The problem is formulated as an optimal control problem with respect to fuel consumption and trip duration, and solved using a Pontryagin minimum principle (PMP) algorithm for a path search problem, such that computations can be carried out online, in real--time. The developed eco–-driving assistance system (EDAS) provides a velocity profile and a sequence of driving modes (and gears) recommendation to the driver, without actively controlling the HDT (human in the loop) and, in practice, allows contextual feedback incorporation from the driver for safety. Simulation results show that the developed methodology is able to provide a velocity profile for a complete route based on known road events and slope information while satisfying all truck operational constraints.
\end{abstract}

\section{Introduction}
In a continuous effort to reduce CO$_2$ emissions, advanced driver or eco--driving assistance systems (AD/EDAS) for heavy--duty vehicles (HDV)---which are responsible for about one quarter of the road transport CO$_2$ emissions in the European Union \cite{euemissionswebsite}---have been in the core of recent technology developments for fossil fuels consumption reduction \cite{edassimulator, sciarretta, PAREDES2019556, ZHU2019562,gao2019evaluation,lee2020model,yutao2020:MPC,wingelaar:2021}. 
Besides, an increased fuel saving also provides a financial incentive, for both internal combustion engines (ICE) and electrical or hybrid motors (longer trips before recharging batteries).

Eco--driving (ED)---see the seminal survey paper \cite{sciarretta}---is a driving strategy that aims to minimize fuel consumption (and, as consequence, emissions) by determining an optimal velocity profile, which can be defined as an optimal control problem (OCP). Strategies that pursue the direct control of engine and service breaks torques, and even gear positions, can be very powerful in the sense of fine control of the HDV dynamics, specially if considering active feedback in the loop; see, e.g., \cite{sciarretta, hellstrom, saerens, SCHORI2013109, powah}. On the other hand, difficulties for real--time implementation of such strategies include safety issues, additional actuator costs and computational time of such optimization problems. Hence, this strategy is mostly applied in driving scenarios where minimum operation is required and automated driving is currently possible, such as adaptive/predictive cruising control (A/PCC) \cite{chen2018real,yutao2020:MPC}. Conversely, EDAS works as an adviser for the driver, which in turn decides to apply or not the advice based on contextual feedback---accounting for safety and no powertrain modifications---, such that one has a driver-in-the loop (DIL) control framework \cite{edassimulator, yutao2020:MPC, nazar, complexsource}. These advice can be delivered utilizing speech, visual interfaces and haptic technology.

Recent studies \cite{nazar,yutao2020:MPC,wingelaar:2021} in which the low--level inputs OCP are abstracted into a driving modes (corresponding to actual operation points of the truck) problem have shown promising results for fuel savings and reduced computational complexity (in the order of less than 1 second for a given segment). This approach allows the computation of the eco--driving advice in real--time, triggered by data--driven events. In these works, optimization was carried either by implementation of Pontryagin minimum principle (PMP) or an event--triggered shrinking horizon model predictive control (MPC).
These works mainly improved the truck longitudinal model and enlarged the set of possible control inputs, but they all considered only deceleration scenarios in a flat---or negligible slope---road. 

In long--hauling situations the starting and ending points of the trip are usually known beforehand, and much of the route information can be obtained before the journey starts. In these cases, slope information can also be previewed and it can be pre--processed in order to obtain uphill, downhill and negligible slope sections (like PCC). 

In this paper we extend the driving modes ED strategy to include slope preview in its solution, while still solving the corresponding hybrid optimization problem via a PMP \cite{Dmitruk:2008,Vinter:2013} approach. As a consequence of inclusion of road slopes in the control problem, addition of two other operational modes are required, namely: acceleration mode and downhill mode. Moreover, because of the addition of the acceleration mode, the developed EDAS is actually able to provide recommendations for the complete route instead of just a deceleration event. Hence, in this work we design a fully operational EDAS that can provide real--time optimized velocity profiles and driving modes advice to the HDT driver.

\section{Heavy--duty truck modeling}
In this section we introduce the main modeling components of the HDT and the driving modes.

\subsection{Engine and retarder models}
The engine speed (in RPM) as related to the wheel velocity is given by
\begin{equation}\label{eq:omega}
    \omega_e = \begin{cases}
    \cfrac{30i_r i_t(y)}{\pi r_w}v & ~~\text{if the gearbox is engaged}\\
    \omega_{idle} & ~~\text{otherwise}
\end{cases}
\end{equation}
where $r_w$ is the wheel radius, the rear axle transmission ratio is $i_r$ and $i_t(y)$ is the transmission ratio at gear $y$. When the gearbox is not engaged, the engine keeps revolving at a constant speed $\omega_{idle}$.

\begin{figure}[htb]
    \centering
    \includegraphics[width=0.95\columnwidth]{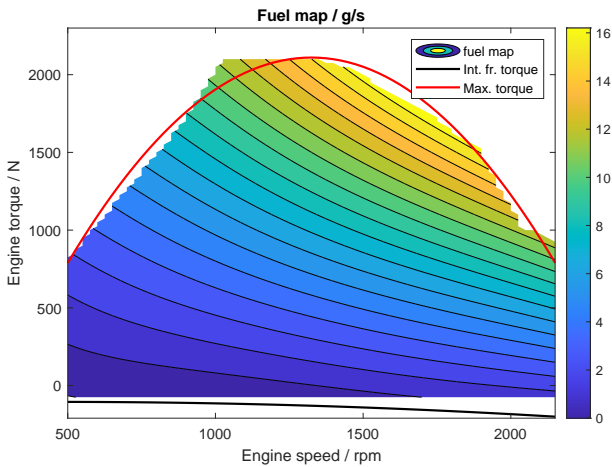}
    \caption{Fuel map as used in \cite{yutao2020:MPC}.}
    \label{fig:fuel_map}
\end{figure}

The engine fuel map (in g/s) is the same as in \cite{yutao2020:MPC} and the points are fitted with a second--order bivariate polynomial given by:
\begin{equation}\label{eq:mf}
\dot{m}_f = \sum_{i=0}^{2}\beta_{i0}\omega_e^i +\sum_{j=1}^{2}\beta_{0j}T_{e}^j + \beta_{11}\omega_e T_e.
\end{equation}
The fitted surface is portrayed in Fig. \ref{fig:fuel_map}. The corresponding maximum engine torque $T_{e,m}$ and internal friction torque $T_{fr}$ (also called engine drag torque) are modeled as
\begin{align}
\label{eq:te_max} T_{e,m} & = t_{e_0} + t_{e_1}\omega_e+ t_{e_2}\omega_e^2 \\
\label{eq:tfric}  T_{fr} & = a_{0} + a_{1}\omega_e+ a_{2}\omega_e^2.
\end{align}

In order to provide additional breaking force, HDTs are equipped with continuous service--breaks systems installed in the vehicle's drivetrain. They convert some of the truck's kinetic energy into heat. The amount of braking torque they can provide varies with the engine speed and its maximum braking torque $T_{eb,m}$ is approximated by the following polynomial
\begin{equation}\label{eq:teb_max}
T_{eb,m} = \frac{t_{b_0}}{\omega_e}+ t_{b_1}+ t_{b_2}\omega_e.
\end{equation}

\subsection{Longitudinal dynamics}
Since in our problem we are also interested in including road slope preview and this information is usually obtained in distance samples just like regular traffic events (like red lights, stop signs), let us first introduce the transformation from time--domain to space--domain:
\begin{equation}
\frac{\dd v}{\dd s} = \frac{\dd v}{\dd t}\frac{\dd t}{\dd s} = \frac{\dd v}{\dd t}\frac{1}{v}. \end{equation}

Thus the truck longitudinal dynamics can be modeled as
\begin{equation}\label{eq:full_dyn}
    \frac{\dd v}{\dd s} = \frac{1}{\left(m+\frac{J_{pt}(y)}{r_w^2}\right)v}\left(\frac{i_r i_t(y)}{r_w}(\eta(T_e-T_{fr})-T_{eb})-F_r\right),
\end{equation}
where $m$ is total mass of the vehicle, $J_{pt}(y)$ is the equivalent inertia of the rotational parts of the truck at gear $y$, and $\eta$ is the combined engine and gear efficiency. The total resistance force (rolling, gravitational and aerodynamic forces) $F_r$ is given by
\begin{equation}\label{eq:resistance}
    F_r = mg(C_r\cos\alpha+\sin\alpha)+\frac{\rho C_{dA}}{2}v^2,
\end{equation}
in which $C_r$ is the rolling coefficient, $\alpha$ is the road slope, $\rho$ is the air density and $C_{dA}$ the combined air drag coefficient times the truck frontal area.

Optimal control problems based directly on \eqref{eq:full_dyn} have the engine torque $T_e$ and the retarder engine braking torque $T_{eb}$ as continuous control inputs. Coupled with the integer gear decision variables $y$, these problems become automatically a \textit{mixed integer} non--linear OCP. This type of HOCP is inherently hard to solve and in many cases are not computationally feasible for real--time applications.

Besides, actual driver behavior imposes some additional deadlock input constraint. For instance, $T_e$ and $T_{eb}$ should not be active at the same time. Also, if $y=0$, then $T_{eb}$ should not be active and $T_e$ is a constant; at the same time, this constant $T_e$ should not produce any propulsion torque in the model. These constraints combined actually allow us to describe the driver and the truck behaviors into specific \textit{driving modes}.

\subsection{Driving modes}
In this section we describe 6 different driving modes that cover, at least, all the driving behaviors required for the solution of our problem. 
Let $i\in\{1,\ldots,6\}=S_M$ be the set of driving modes and $y\in\{0,\ldots,12\}=S_G$ be the set of gears, where $y=0$ means neutral. We then define the driving mode $q$ as $q:\mathbb{N}_{\geq 1}\to S_M\times S_G=:\mathbb{Q}$.

Each of these driving modes represent one realistic configuration of the truck (without any addition of further binary selection inputs) such that each one has its own fixed version of \eqref{eq:full_dyn} described by $f_{\qiy}$.
The translation of the possible dynamics into given driving modes is one step further into the problem type simplification, since for each mode the continuous input profile is fixed, which will render the problem a path search problem between the driving modes and their respective gear.

\subsubsection{Cruising mode}
In cruising mode the engine provides only enough power to keep the truck at a constant velocity. This implies that
\begin{equation}\label{eq:dyn_cruise}
    f_{q} = 0.
\end{equation}

The requested engine torque to keep the truck at a constant velocity is computed such that
\begin{equation}\label{eq:te_cruising}
    T_{e} = \frac{r_w}{i_r i_t(y)\eta}F_r + T_{fr}\leq T_{e,m},
\end{equation}
and the fuel consumption is then given by $\dot{m}_f(\omega_e,T_e)$.

Notice that because we are considering slope information, cruising itself (i.e., constant velocity driving) can be achieved in different ways. Thus, whenever we refer to \textit{cruising} as a \textit{mode}, this will imply that the engine needs to provide power to keep constant velocity.

\subsubsection{Eco--rolling mode}
Eco--roll mode is a special case of the gearbox state. In this case, the gear is set to neutral ($y=0$) and the HDT glides only on the resistance forces. Its dynamics are expressed by
\begin{equation}\label{eq:dyn_eco}
    f_{q} = \frac{-F_r}{\left(m+\frac{J_{pt}(0)}{r_w^2}\right)v}.
\end{equation}

In this mode the engine operates at a constant engine speed $\omega_{idle}$ and constant engine torque leading to a constant fuel consumption $\dot{m}_f=m_{f_{idle}}$.

\subsubsection{Coasting mode}
In coasting mode, the gearbox is engaged in one of the gears $y>0$, such that the internal engine friction adds another resistant torque. This is expressed by
\begin{equation}\label{eq:dyn_coast}
    f_{q} = \frac{-1}{\left(m+\frac{J_{pt}(y)}{r_w^2}\right)v}\left(\frac{i_r i_t(y)}{r_w}\eta T_{fr}+F_r\right).
\end{equation}
In this case, the engine revolves based on the wheel speed, so there is no fuel consumption, i.e.,  $\dot{m}_f=0$

\subsubsection{Engine breaking mode}
In this mode, we consider that the retarder operates in its maximum torque line \eqref{eq:teb_max} to deliver enough breaking force to the vehicle, such that the state dynamic is given by
\begin{equation}\label{eq:dyn_eb}
    f_{q} = \frac{-1}{\left(m+\frac{J_{pt}(y)}{r_w^2}\right)v}\left(\frac{i_r i_t(y)}{r_w}(\eta T_{fr}+T_{eb,m})+F_r\right).
\end{equation}
Moreover, because the retarder is coupled to the gearbox, and the engine revolves based on the wheel speed, then there is no fuel consumption $\dot{m}_f=0$.

\subsubsection{Downhill mode}
This is a special case of the engine breaking mode and it only applies when $F_r<0$. In this case, the available retarder torque, but not necessarily its maximum, is used to keep the vehicle in a constant velocity while going downhill, that is, 
\begin{equation}\label{eq:dyn_downhill}
    f_{q} = 0.
\end{equation}

In order to keep this constant velocity, the engine break torque is computed such that
\begin{equation}\label{eq:teb_downhill}
    T_{eb} = -\frac{r_w}{i_r i_t(y)}F_r -\eta T_{fr}\leq T_{eb,m}.
\end{equation}
Finally, there is also no fuel consumption $\dot{m}_f=0$.

\subsubsection{Maximum torque acceleration mode}
Even though acceleration can be achieved at many different engine torque points, in order to fix it as a driving mode, we will assume that it is done using the maximum torque line, i.e., $T_e=T_{e,m}$ in \eqref{eq:te_max}. This is suggested since full throttle acceleration is considered to be the most efficient way to gain speed \cite{stoicescu:1995:maxtorque}. Thus the longitudinal dynamics take the form
\begin{equation}\label{eq:dyn_acc}
    f_{q}= \frac{1}{\left(m+\frac{J_{pt}(y)}{r_w^2}\right)v}\left(\frac{i_r i_t(y)}{r_w}(\eta(T_{e,m}-T_{fr}))-F_r\right),
\end{equation}
and the fuel consumption is then given by $\dot{m}_f(\omega_e,T_{e,m})$.

\subsection{Driveability and comfort constraints}
For the envisioned EDAS advice to be well--received by the drivers and put to practice, then some practical driveability---the consumer perception of the vehicle---and comfort constraints (which can be very subjective) must also be taken into account.
%Driveability is the consumer perception of the vehicle. How well the vehicle responds (movement, sound) to the driver expectations affects the driver's emotion and opinions of how well the truck drives. The driver has few inputs to the vehicle to convey their intentions: Accelerator Pedal, Brake Pedal, Steering Wheel, and Transmission Mode. A group of drivers would have different expectations and give different subjective ratings on the shift quality and shift performance for the same vehicle.
In this work, the main constraint will be placed on the maximum acceleration and deceleration of the truck (notice that $vf_{\qiy}$ gives acceleration in m/s$^2$). 

These constraints usually depend on the type of driving: for example, maximum acceleration and jerk are analyzed for a Pulse and Glide behavior in \cite{sohn:2019:driveability}, whereas in \cite{bae:2019:comfortable} acceleration is analyzed for comfort riding in a shuttle bus (which may include standing--up passengers), yielding $a_{max} = 1.2$ m/s$^2$, and $a_{min} = -0.7$ m/s$^2$. On the other hand, uphill scenarios may require additional acceleration. Thus, in this paper we imposed the following constraint
\begin{equation}\label{eq:a_maxmin}
    a_{max} = -a_{min} = 2~\text{m/s}^2.
\end{equation}

The formulation with driving modes also requires some attention: a solution that requires frequent mode switching may also not be practical. More specifically, we noticed a frequent switching from eco--rolling to coasting mode and back, when the truck is going down a hill while already at the maximum legal velocity. In order to avoid this behavior, we \textit{remove} eco--rolling mode from the search if all three conditions below hold
\begin{enumerate}
    \item $\text{sign}(F_r)=-\text{sign}(f_{q_2})$;
    \item $(v_{lim}-v)<1.5$~km/h;
    \item previous mode not eco--rolling.
\end{enumerate}

\section{Optimal control problem formulation}\label{sec:ocp}
The main goal in a eco--driving solution is to reduce the fuel consumption over the journey. However, if this is the only cost being minimized, then the optimal solution is always driving in low velocities. Or in the most degenerating case, not moving at all. Practical applications, especially in delivery situations, also penalize the trip duration. Hence, we formulate the cost function as
\begin{equation}\label{eq:cost_function}
    J =  \int_{s_0}^{s_f} \underbrace{\left(W_1\frac{\dot{m}_f(\qiy)}{v} + W_2\frac{1}{v}\right)}_{g_{\qiy}(v)}  \dd s,
\end{equation}
where $W_1$ and $W_2$ are fuel consumption and trip duration weights, respectively. Let $J$ represents the mechanical work developed by the truck. This means that the integrand $g_{\qiy}(v)$ has units of newtons $N$, $W_1$ is $m^2/s^2$ and $W_2$ is $kgm^2/s^3$. Thus the ratio $W_2/W_1$ is $kg/s$ which is a measure of the average fuel consumption per distance.

Hence, the OCP can be formulated as 
\begin{equation}\label{eq:ocp}
\begin{aligned}
    \min_{\{q^j\}_{{j\in\mathbb{N}_{[1,N]}}}} & ~~ J \\
    \text{subject to} & ~~ \frac{\dd v}{\dd s}=f_{\qiy}\\
    & ~~ v_{min}\leq v\leq v_{lim}\\
    & ~~ a_{min}\leq\dot{v}\leq a_{max}\\
    & ~~ \omega_{min}\leq\omega\leq\omega_{max}\\
    & ~~ 0<T_e<T_{e,m}\\
    & ~~ 0<T_{eb}<T_{eb,m}\\
    & ~~ v(s_0)=v_0 ~\text{and}~ v(s_f)=v_f.
\end{aligned}
\end{equation}
Notice that problem \eqref{eq:ocp} is slightly reduced to a mode search over the feasible driving mode sequence, i.e, for $\qiy\in\mathbb{Q}$.

One way to solve problem \eqref{eq:ocp} is via Pontryagin minimum principle (PMP) \cite{Dmitruk:2008,Vinter:2013,nazar,wingelaar:2021}, if constraints along the path are omitted, i.e., when only the two--point boundary (TPB) constraints are considered. For that, we define the Hamiltonian for each subsystem $\qiy$ associated to the control problem as
\begin{equation}\label{eq:hamilt}
H_{q} = g_{\qiy}(v) + \lambda f_{\qiy},    
\end{equation}
where $g_{\qiy}(v)$ is defined in \eqref{eq:cost_function}, $f_{\qiy}$ is one of \eqref{eq:dyn_cruise}, \eqref{eq:dyn_eco}, \eqref{eq:dyn_coast}, \eqref{eq:dyn_eb}, \eqref{eq:dyn_downhill}, \eqref{eq:dyn_acc}, and the variable $\lambda$ is known as costate \cite{Dmitruk:2008,Vinter:2013,VANKEULEN2014:ineqPMP}. Notice that we utilized a simplified version of the Hamiltonian which does not include costate for the inequality constraints \cite{VANKEULEN2014:ineqPMP}. The motivation for this is better understood in sections \ref{ssec:DHOCP} and \ref{sec:issues}, where we assume a discretized version of problem \eqref{eq:ocp}.

The evolution of the state vector follows
\begin{equation}\label{eq:vdot}
\frac{\dd v}{\dd s} = \frac{\partial H_{q}}{\partial \lambda} =  f_{\qiy},
\end{equation}
while the costate dynamics is given by
%\begin{equation}\label{eq:lambda_dot}
%\begin{aligned}
%\frac{\dd \lambda}{\dd s} & = -\frac{\partial H_{q}}{\partial v} \\
%& = -\frac{W_1\dot{m}_f}{v^2} + \frac{W_1}{v}\frac{\partial\dot{m}_f}{\partial %v} - \frac{W_2}{v^2}+ \lambda\frac{\partial f_{\qiy}}{\partial v}    
%\end{aligned}
%\end{equation}
\begin{equation}\label{eq:lambda_dot}
\frac{\dd \lambda}{\dd s}  = -\frac{\partial H_{q}}{\partial v}
= -\frac{W_1\dot{m}_f}{v^2} + \frac{W_1}{v}\frac{\partial\dot{m}_f}{\partial v} - \frac{W_2}{v^2}+ \lambda\frac{\partial f_{\qiy}}{\partial v}    
\end{equation}
where $\frac{\partial\dot{m}_f}{\partial v}$ can be computed by the chain rule
\begin{equation}\label{eq:dmf_dv}
\frac{\partial\dot{m}_f}{\partial v} = \frac{\partial\dot{m}_f}{\partial \omega_e}\frac{\partial \omega_e}{\partial v} + \frac{\partial\dot{m}_f}{\partial T_e}\frac{\partial T_e}{\partial v}.
\end{equation}
Notice that the right--hand side of \eqref{eq:dmf_dv} is easily computable via \eqref{eq:omega}, \eqref{eq:mf}, \eqref{eq:te_max} and \eqref{eq:te_cruising}. 

Since every state dynamics $f_{\qiy}$ different from zero is dependent on the the current state value and no input acts directly in the state dynamics, it is fair to assume velocity continuity at mode switching point, such that there is no jumps in the velocity profile. Furthermore, because of continuity of $v$, we shall also assume continuity of $\lambda$, even when $\qiy$ changes.

\subsection{Discrete HOCP}\label{ssec:DHOCP}
In order to further simplify the computational complexity of the OCP, the problem is discretized into $N$ distance samples of length $\Delta s=\frac{s_f-s_0}{N}$ on which it is assumed the driving mode $\qiy$ is piecewise constant on each interval $s \in (s_k, s_{k+1}], k = 0,\ldots, N$ \cite{nazar,wingelaar:2021}. The cost function then becomes
\begin{equation}
    J_d =  \sum_{k=0}^{N-1} g_{\qiy}^k(v_{k+1})  \Delta s \\
\end{equation}
and the state and costate updates are given by
\begin{align}
\label{eq:v_update} v_k & = v_{k+1} - \frac{\partial H_{q}^{k+1}}{\partial\lambda}  \Delta s, \\
\label{eq:l_update} \lambda_k & = \lambda_{k+1} + \frac{\partial H_{q}^{k+1}}{\partial v}\Delta s, \\
H_{q}^{k+1} & = g_{\qiy}^k(v_{k+1}) + \lambda_{k+1} f_{\qiy}.  
\end{align}
Notice that with the discretization of the road segment, which implies discretization of the state dynamics, and the enumeration of the possible driving modes, our problem is transformed into path search problem such that constraint violation can be dealt with by simply removing the node from the path search.

The optimal $q$ on distance sample $s_k$ is found by:
\begin{equation}
q^{k*} = \underset{q\in\mathbb{Q}}{\arg\min}
%\left(
%\underset{y\in\mathbb{Y}}{\arg\min}
~H_{q}^{k+1}(v_{k+1},\lambda_{k+1}),%\right)
\end{equation}
that is, at each distance sample $s_k$, we compute the Hamiltonian for all subsystems (i.e., modes and respective gear), and select the mode/gear that yields the minimum of them all.
\subsection{Implementation issues}\label{sec:issues}
\subsubsection{Forward Euler discretization}
In previous works \cite{nazar,wingelaar:2021} direct application of the forward Euler forms \eqref{eq:v_update} and \eqref{eq:l_update} were applied. However, in these same works, actual slope data was not considered. In order to improve numerical stability of the algorithm due to these variations on the slope data, updates \eqref{eq:v_update} and \eqref{eq:l_update} are performed using a Runge--Kutta type 4 integration method \cite{butcher2016numerical}.

\subsubsection{Minimum velocity and convergence}
Notice that the velocity derivatives for every mode depend on the inverse of the current velocity. This imposes an undefined dynamics if the truck reaches standstill position, i.e., when $v=0$. In order to avoid this issue, we imposed the following constraint
\begin{equation}
    v_{min} = 8~\text{km/h}.
\end{equation}
This means that standstill velocity events, like red lights, are also treated as $8$~km/h, and we assume the truck reaches $0$~km/h by its own inertia or short application of regular brakes.

Furthermore, since we solve the problem backwards in time, convergence to the initial velocity constraint $v_0$ is assessed via two error threshold of different magnitudes. We say the optimization has converged:
\begin{itemize}
    \item if for every new $\lambda_N^j$, the error $|v_0-v_0^*|\leq 0.01$~km/h; or
    \item if $|\lambda_N^j - \lambda_N^{j-1}|\leq 0.0002$, the error $|v_0-v_0^*|\leq 1$~km/h.
\end{itemize}
The second condition is useful to avoid new iterations, and thus longer optimization time, when the costate (and thus the optimized velocity profile) is only marginally changing.

\subsubsection{Costate initialization}
Note that the choice of $\lambda_N$ fully determines the state trajectory; however the value of $\lambda_N$ that leads to $v_0$ is initially unknown. 
Costate initialization in TPB value problems is a fundamental issue in shooting methods (and thus in PMP) \cite{sciarretta}, and there is no formal method that can address the initialization problem. For this work we applied the same approach in \cite{wingelaar:2021} where a bi--section search scheme based on the error sign is used.

\subsubsection{Path constraints}
In order to still apply PMP, path constraints are dealt with indirectly, as follows.
During the backwards generation of the velocity profile and costate trajectory, path constraints are checked and if a violation is detected, the corresponding mode is not included in the Hamiltonian evaluation. In case no combination of mode/gear satisfy all the constraints at a given sample, either acceleration or cruising mode is selected in order to force an incorrect $v_0^*$, so $\lambda_N$ is updated in the next iteration. If any solution can be found, then no advice is given.

\section{Route information}
In long--hauling situations the starting and ending points of the trip are usually known beforehand, and much of the route information can be obtained \textit{a priori}. For example, static road events---and where they happen---like red lights, stop signs and changing in speed limit are easily available. 

These static road events can be used to define the TPB value problem, where the initial velocity is the current HDT velocity and the final velocity is defined by the event type. 

Furthermore, slope information can also be obtained and it can be pre--processed in order to obtain uphill, downhill and negligible slope sections. Determining uphill and downhill sections beforehand is very much alike what PCC---which is already installed in many HDT nowadays---does, so this also makes the information feel more ``at home" for the driver. It also helps finding places where one can use the slope to accelerate or prepare for a steep uphill.

%From the point of view of the algorithm, the negligible slope are sections that do not exceed $\pm5$~m height.

In this way, instead of finding the optimal velocity profile for the complete route at once, the profiles are obtained for each segment. This allows a procedure where, if the driver decides not to follow an advice for the current segment, we only need to recompute the next one.

\begin{figure}[tb]
    \centering
    \includegraphics[width=0.98\columnwidth]{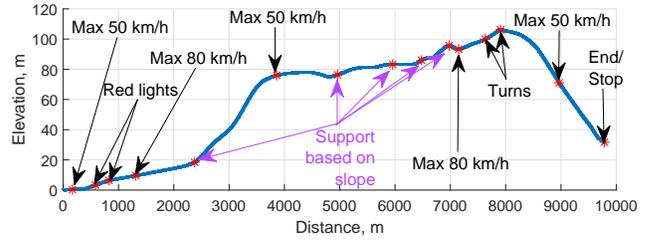}
    \caption{Route segmentation based on static road events and slope information (negligible slopes and uphill parts, mainly).}
\label{fig:elevation_maas}
\end{figure}

An example of the output of such pre--processing is presented in Fig. \ref{fig:elevation_maas} for a small route going from the United World College of Maastricht, Maastricht, The Netherlands to the entrance of Valkenburg, the Netherlands, via N590. The elevation profile is relative to the initial position, and changes in maximum speed tells the limit from that point onward.

% I'll just state main functionalities:
%\begin{itemize}
%    \item Downhill section: computes, when possible, the minimum speed at the beginning of the section such that if the truck eco--rolls until the end it reaches maximum legal speed. This minimum speed is limited to $v_{lim}-10$ (PCC compliant, info obtained from Adi)
%    \item Uphill sections: set the event velocity depending on maximum acceleration/torque for a given maximum slope.
%\end{itemize}

\section{Illustrative example}
In this section we provide a simulation result of our proposed methodology using real route information for the route presented in Fig. \ref{fig:elevation_maas}. For the simulation, the truck parameters are given in Table \ref{tab:truck} and the engine speed is limited between $\omega_{min} = 550$~rpm and $\omega_{max} = 2200$~rpm. 

\begin{table}[tb]
\caption{Truck parameters used in the simulations, as in \cite{yutao2020:MPC}.}
\centering
\begin{tabular}{cc|cc|cc}
Param.   & Value  & Param.   & Value  & Param.               & Value               \\ \hline
$m$      & 30e3   & $i_t(1)$ & 15.86  & $i_t(9)$             & 2.1                 \\
$C_r$    & 0.009  & $i_t(2)$ & 12.33 & $i_t(10)$            & 1.63                \\
$C_{dA}$ & 6.24   & $i_t(3)$ & 9.57   & $i_t(11)$            & 1.29                \\
$r_w$    & 0.492  & $i_t(4)$ & 7.44   & $i_t(12)$            & 1                   \\
$i_r$    & 2.6875 & $i_t(5)$ & 5.87   & $J_{pt}(y)$          & 83.8+19.56$i_t^2$   \\
$g$      & 9.806  & $i_t(6)$ & 4.57   & $\omega_{idle}$      & 550                 \\
$\rho$   & 1.205  & $i_t(7)$ & 3.47   & $T_{e_{idle}}$       & 150                 \\
$\eta$    & 0.98  & $i_t(8)$ & 2.7    & $\dot{m}_{f_{idle}}$ & 0.27                \\ \hline
$t_{e_0}$ & -1298 & $t_{e_1}$ & 5.144 & $t_{e_2}$ & -1.941e-3 \\
$t_{b_0}$ & -4.198e6 & $t_{b_1}$ & 6961.432 & $t_{b_2}$ & -1.581\\
$a_{0}$ & 112.5 & $a_{1}$ & -0.0314 & $a_{2}$ & 3.36e-5 \\ \hline
\end{tabular}
\label{tab:truck}
\end{table}

In this example we solve the OCP for each segment of the route in Fig. \ref{fig:elevation_maas}. For red lights we always assume the worst case scenario, such that the driver needs to stop. Notice that since this is an advice, if the driver sees that the red light is actually green as he approaches it, they can decide to overwrite the advice (for example, by pressing the throttle pedal again). 
We first provide a simple comparison with a solution using the continuous input model, and solved using CasADi/IPOPT, just like in \cite{yutao2020:MPC}. We then provide the solution for the full route using our PMP approach.

\subsection{Comparison with continuous inputs}
\begin{figure}[htb]
    \centering
    \includegraphics[width=0.98\columnwidth]{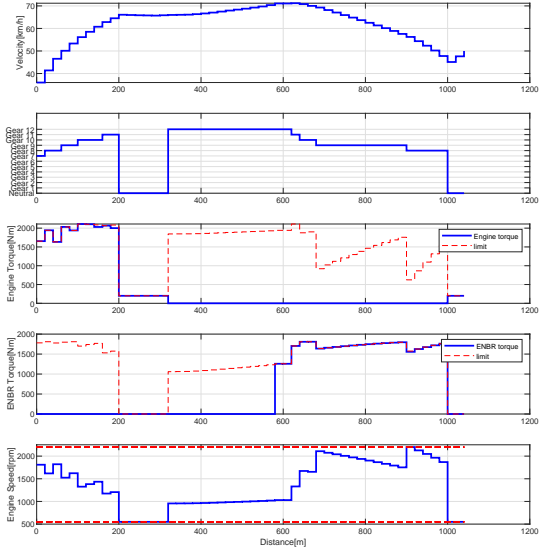}
    \caption{Optimized velocity and gearshift profiles for the solution using the continuous-time inputs.}
    \label{fig:casadi20}
\end{figure}

\begin{figure}[htb]
    \centering
    \includegraphics[width=0.98\columnwidth]{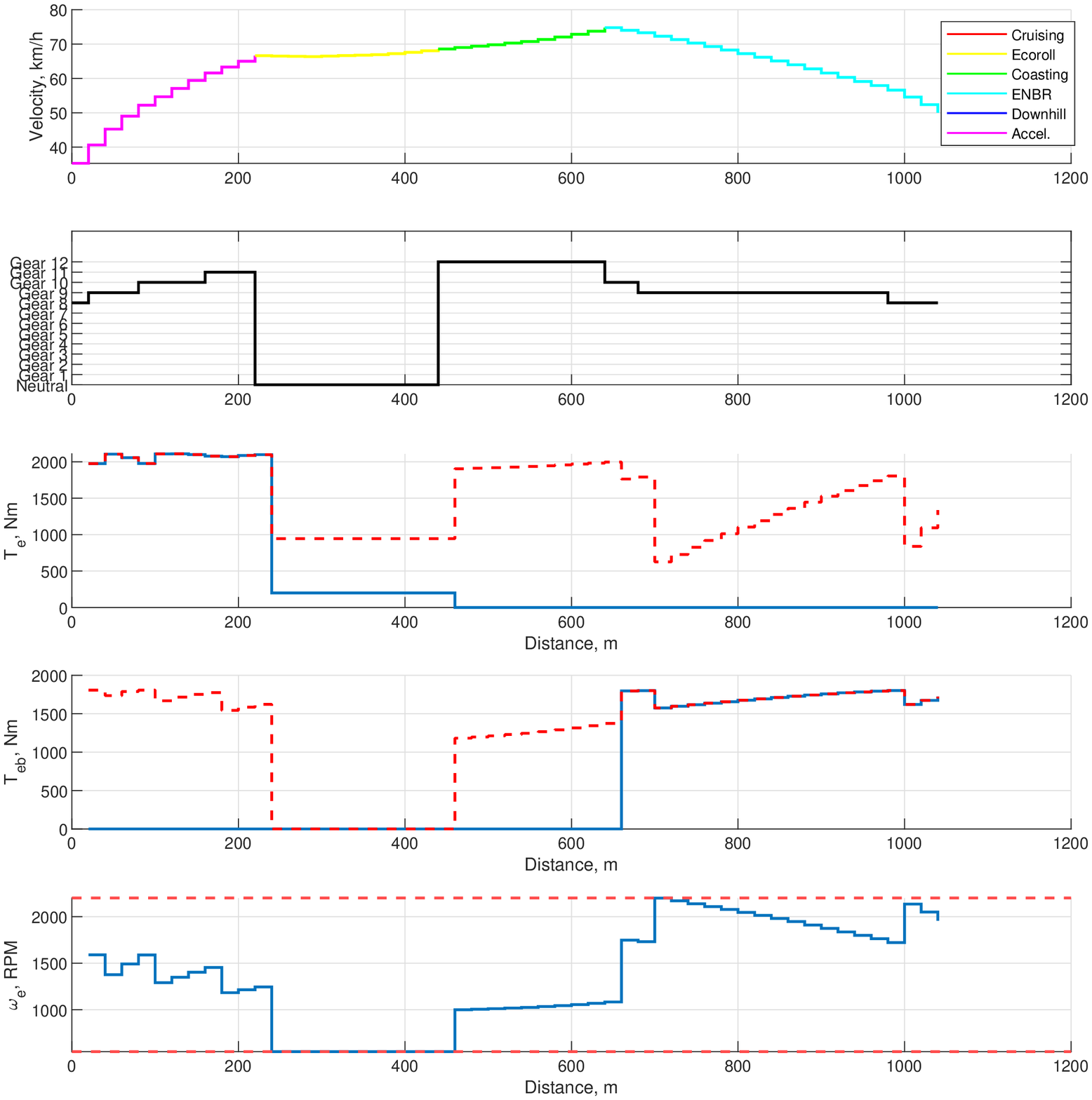}
    \caption{Optimized velocity and gearshift profiles, and driving modes advice for the solution with the PMP approach.}
    \label{fig:pmp20}
\end{figure}

For the comparison we chose to solve for the segment 14 (between 8 and 9 km approximately), where the truck leaves a turn at 36 km/h and should reach a speed of 50 km/h, while the maximum speed in the segment is 80 km/h. Furthermore, the slope distance points are rounded to a meter and then linearly interpolated, while the discretized interval is $\Delta s = 20$~m. The obtained solutions are presented in figs. \ref{fig:casadi20} and \ref{fig:pmp20} for the IPOPT and the PMP solvers respectively.


Notice that the achieved velocity profile for both cases are very similar, even following the same state of the engine, with a small deviation at the when the the solution with IPOPT accelerates again. However, computational time for the first case was about 7 minutes with a cost function of $J=2136$, while for PMP was 0.5 seconds with $J=2011$. Although the difference in cost function value is non--significant and may differ only due to numerical integration, the computational time difference is large and shows that our approach is deemed to real--time implementation.

\subsection{Full route solution with PMP}
\begin{figure}[htb]
    \centering
    \includegraphics[width=0.98\columnwidth]{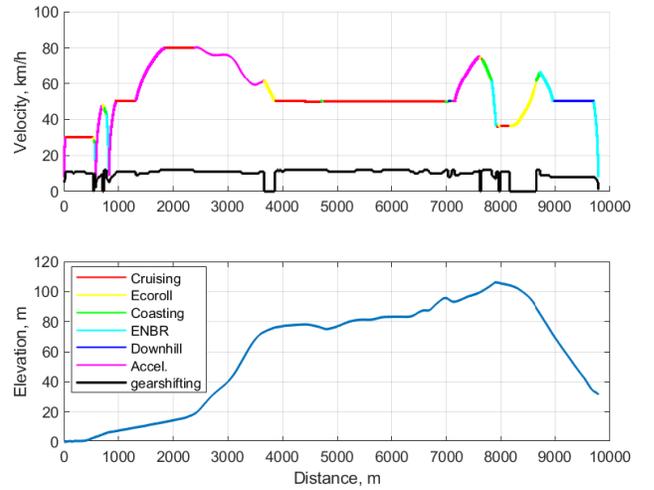}
    \caption{Optimized velocity and gearshift profiles, and driving modes sequence. The legend on the elevation plot is with respect to the velocity profile and gear shift plot; it was placed on the lower plot due to space constraints.}
    \label{fig:v_maas}
\end{figure}

For the complete solution of the route, we have first reduced the discretized interval to $\Delta s = 1$~m, allowing better control of the driving modes changes. The combined velocity profile for each of the 15 segments and the respective driving mode is shown in Fig. \ref{fig:v_maas}.
The computed fuel consumption for this route is 4.018~kg ($\approx$ 4.727~L) with an estimated trip duration of 11.5 minutes.

Notice that, whenever possible, the optimized profile leads to accelerating to the maximum legal velocity, which is coherent with penalizing the trip duration. Besides, deceleration is always possible with one of the decelerating modes, so no disc--breaking is required. Furthermore, for approximately the last $2$~km, almost no fuel is used, only for idling in eco--roll while gaining speed using the slope.

\begin{table}[tb]
\caption{Computational time, in seconds, for each segment of the Maastricht route.}
\setlength{\tabcolsep}{3pt}
\centering
\begin{tabular}{cc|cc|cc|cc|cc}
segm. & time & segm. & time & segm. & time & segm. & time & segm. & time \\ \hline
1     & 0.40 & 4     & 0.74 & 7     & 0.17 & 10    & 0.02 & 13    & 0.41 \\
2     & 0.49 & 5     & 1.29 & 8     & 0.04 & 11    & 0.06 & 14    & 1.06 \\
3     & 0.40 & 6     & 1.45 & 9     & 0.02 & 12    & 0.81 & 15    & 0.88
\end{tabular}
\label{tab:comp_time}
\end{table}

In Table \ref{tab:comp_time} we show the computational time of the optimization in each segment, for the code running in MATLAB$^{\tiny{\text{\textregistered}}}$ in a Intel$^{\tiny{\text{\textregistered}}}$ Core$^{\tiny{\text{\texttrademark}}}$ i7-9750H CPU @ 2.60GHz. In \cite{yutao2020:MPC}, where only the deceleration scenario in a flat road is considered, computational time for a nonlinear MPC approach by direct torque optimization is on the order of $80.4$~s. On the other hand, for the approach with driving modes, they obtained a computational time of $0.33$~s using the IPOPT optimization toolbox with a much coarse grid of $\Delta s = 20$~m. In our case, with real slope data and a finer grid, the computation time stays below the mark of 2 seconds for every segment, which is already applicable in real--time. If further we consider the implementation in a dedicated ECU running a C/C++ code, it is expected this time can drop by at least 10 times \cite{nazarmaster}.

\begin{figure}[htb]
    \centering
    \includegraphics[width=0.98\columnwidth]{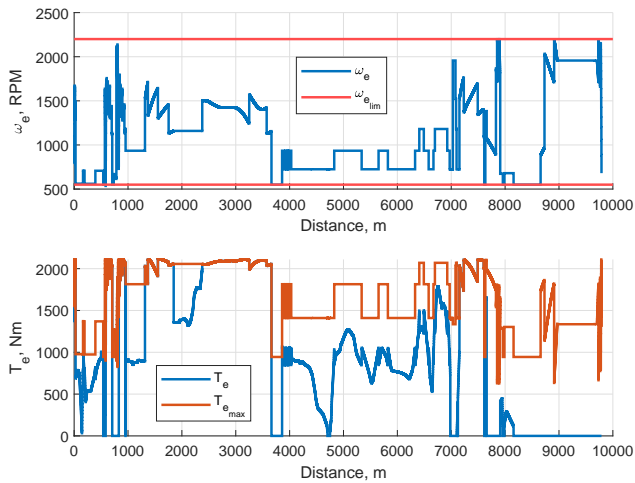}
    \caption{Engine speed and engine torque as a by--product of the optimization. The red lines are the engine limits and for the torque, the limits are given as a function of the engine speed.}
    \label{fig:engine_maas}
\end{figure}

It is worth to mention that it is also possible to achieve real--time feasible solutions to the eco--driving problem by analytic methods for solving the OCP without driving modes, as done in \cite{sciarretta}. In fact, these methods, which involve certain problem specific simplifications/assumptions, could be adapted to our OCP problem with driving modes in order to obtain a fast estimate of the initial costate/optimal velocity trajectory.

%\subsection{Paris to Rouen}
%\hl{I'm thinking if I should/could include one of those more complete examples from the data I got from IFPEN that include even traffic events... but then HOW do I explain it without citing them? So, for the space that's left a have two options: either add this new example, or plot engine speed and torque for the previous example and use the remaining space to explain better some other part of the paper. }

\section{Conclusions}\label{sec:5}
In this paper, a complete eco--driving strategy based on a finite number of driving modes for HDTs in long--hauling scenarios was developed to cope with different route events and also real road slope data preview. The developed EDAS allows contextual feedback incorporation from the driver because it only provides a recommendation of driving modes, that can be computed in real--time. Simulation results showed that the developed methodology was able to provide a velocity profile for a complete route based on known road events and slope information while satisfying all truck operational constraints.

It is of interest to further validate the developed eco-–driving methodology by using the VECTO tool \cite{euvecto} and further comparison with the corresponding OCP without the driving modes. Inclusion of real-time information such as current traffic, red light state, and preceding vehicle can also be considered.

% Acknowledgments---Will not appear in anonymized version
\section*{Acknowledgments}
The authors are grateful to the following partners in the LONGRUN EU project for many useful discussions and feedback on our research: John Kessels, Guus Arts, Bram Hakstege (DAF), and Ilias Cheimariotis and Christophe Cornet (IFPEN).

\bibliographystyle{IEEEtran}
\bibliography{references}

\begin{thebibliography}{10}
\providecommand{\url}[1]{#1}
\csname url@samestyle\endcsname
\providecommand{\newblock}{\relax}
\providecommand{\bibinfo}[2]{#2}
\providecommand{\BIBentrySTDinterwordspacing}{\spaceskip=0pt\relax}
\providecommand{\BIBentryALTinterwordstretchfactor}{4}
\providecommand{\BIBentryALTinterwordspacing}{\spaceskip=\fontdimen2\font plus
\BIBentryALTinterwordstretchfactor\fontdimen3\font minus
  \fontdimen4\font\relax}
\providecommand{\BIBforeignlanguage}[2]{{%
\expandafter\ifx\csname l@#1\endcsname\relax
\typeout{** WARNING: IEEEtran.bst: No hyphenation pattern has been}%
\typeout{** loaded for the language `#1'. Using the pattern for}%
\typeout{** the default language instead.}%
\else
\language=\csname l@#1\endcsname
\fi
#2}}
\providecommand{\BIBdecl}{\relax}
\BIBdecl

\bibitem{euemissionswebsite}
{European Environment Agency}, ``Carbon dioxide emissions from {Europe's}
  heavy-duty vehicles,''
  \url{https://www.eea.europa.eu/themes/transport/heavy-duty-vehicles/carbon-dioxide-emissions-europe},
  {A}ccessed: 08-11-2021.

\bibitem{edassimulator}
T.~J. {Daun}, D.~G. {Braun}, C.~{Frank}, S.~{Haug}, and M.~{Lienkamp},
  ``Evaluation of driving behavior and the efficacy of a predictive eco-driving
  assistance system for heavy commercial vehicles in a driving simulator
  experiment,'' in \emph{16th International IEEE Conference on Intelligent
  Transportation Systems}, 2013, pp. 2379--2386.

\bibitem{sciarretta}
A.~Sciarretta, G.~De~Nunzio, and L.~Ojeda, ``Optimal ecodriving control:
  Energy-efficient driving of road vehicles as an optimal control problem,''
  \emph{Control Systems, IEEE}, vol.~35, pp. 71--90, 2015.

\bibitem{PAREDES2019556}
J.~{Flores Paredes}, G.~{Padilla Cazar}, and M.~Donkers,
  ``\BIBforeignlanguage{English}{A shrinking horizon approach to eco-driving
  for electric city buses: implementation and experimental results},'' in
  \emph{\BIBforeignlanguage{English}{9th IFAC International Symposium on
  Advances on Automotive Control}}, no.~5, 2019, pp. 556--561.

\bibitem{ZHU2019562}
J.~Zhu, C.~Ngo, and A.~Sciarretta, ``Real-time optimal eco-driving for
  hybrid-electric vehicles,'' \emph{IFAC-PapersOnLine}, vol.~52, no.~5, pp.
  562--567, 2019.

\bibitem{gao2019evaluation}
Z.~Gao, T.~LaClair, S.~Ou, S.~Huff, G.~Wu, P.~Hao, K.~Boriboonsomsin, and
  M.~Barth, ``Evaluation of electric vehicle component performance over
  eco-driving cycles,'' \emph{Energy}, vol. 172, pp. 823--839, 2019.

\bibitem{lee2020model}
H.~Lee, N.~Kim, and S.~W. Cha, ``Model-based reinforcement learning for
  eco-driving control of electric vehicles,'' \emph{IEEE Access}, vol.~8, pp.
  202\,886--202\,896, 2020.

\bibitem{yutao2020:MPC}
Y.~Chen and M.~Lazar, ``Real-time driving mode advice for eco-driving using
  {MPC},'' \emph{IFAC-PapersOnLine}, vol.~53, no.~2, pp. 13\,830--13\,835,
  2020.

\bibitem{wingelaar:2021}
B.~Wingelaar, G.~R. {Gonçalves da Silva}, M.~Lazar, Y.~Chen, and J.~T. B.~A.
  Kessels, ``Design and assessment of an eco-driving pmp algorithm for optimal
  deceleration and gear shifting in trucks,'' in \emph{2021 IEEE Conference on
  Control Technology and Applications (CCTA)}, San Diego, California, 2021, pp.
  8--13.

\bibitem{hellstrom}
E.~Hellstr{\"o}m, J.~{\AA}slund, and L.~Nielsen, ``Design of an efficient
  algorithm for fuel-optimal look-ahead control,'' \emph{Control Engineering
  Practice}, vol.~18, no.~11, pp. 1318--1327, 2010.

\bibitem{saerens}
B.~Saerens, ``Optimal control based eco-driving,'' Ph.D. dissertation, KU
  Leuven, 2012.

\bibitem{SCHORI2013109}
M.~Schori, T.~J. Boehme, B.~Frank, and M.~Schultalbers, ``Solution of a hybrid
  optimal control problem for a parallel hybrid vehicle,'' \emph{IFAC
  Proceedings Volumes}, vol.~46, no.~21, pp. 109--114, 2013.

\bibitem{powah}
Q.~{Jin}, G.~{Wu}, K.~{Boriboonsomsin}, and M.~J. {Barth}, ``Power-based
  optimal longitudinal control for a connected eco-driving system,'' \emph{IEEE
  Transactions on Intelligent Transportation Systems}, vol.~17, no.~10, pp.
  2900--2910, 2016.

\bibitem{chen2018real}
H.~Chen, L.~Guo, H.~Ding, Y.~Li, and B.~Gao, ``Real-time predictive cruise
  control for eco-driving taking into account traffic constraints,'' \emph{IEEE
  Transactions on Intelligent Transportation Systems}, vol.~20, no.~8, pp.
  2858--2868, 2018.

\bibitem{nazar}
Y.~Chen, N.~Rozkvas, and M.~Lazar, ``Driving mode optimization for hybrid
  trucks using road and traffic preview data,'' \emph{Energies}, vol.~13, 2020.

\bibitem{complexsource}
A.~Jamson, D.~Hibberd, and N.~Merat, ``Interface design considerations for an
  in-vehicle eco-driving assistance system,'' \emph{Transportation Research
  Part C: Emerging Technologies}, vol.~58, 2015.

\bibitem{Dmitruk:2008}
A.~Dmitruk and A.~Kaganovich, ``The hybrid maximum principle is a consequence
  of {Pontryagin Maximum Principle},'' \emph{Systems \& Control Letters},
  vol.~57, no.~11, pp. 964--970, 2008.

\bibitem{Vinter:2013}
R.~B. Vinter, \emph{Optimal Control and Pontryagin's Maximum Principle}.\hskip
  1em plus 0.5em minus 0.4em\relax London: Springer London, 2013, pp. 1--9.

\bibitem{stoicescu:1995:maxtorque}
A.~P. Stoicescu, ``On fuel--optimal velocity control of a motor vehicle,''
  \emph{International Journal of Vehicle Design}, vol.~16, no. 2-3, pp.
  229--256, 1995.

\bibitem{sohn:2019:driveability}
C.~Sohn, J.~Andert, and R.~N.~N. Manfouo, ``A driveability study on automated
  longitudinal vehicle control,'' \emph{IEEE Transactions on Intelligent
  Transportation Systems}, vol.~21, no.~8, pp. 3273--3280, 2019.

\bibitem{bae:2019:comfortable}
I.~Bae, J.~Moon, and J.~Seo, ``Toward a comfortable driving experience for a
  self-driving shuttle bus,'' \emph{Electronics}, vol.~8, no.~9, p. 943, 2019.

\bibitem{VANKEULEN2014:ineqPMP}
T.~{van Keulen}, J.~Gillot, B.~{de Jager}, and M.~Steinbuch, ``Solution for
  state constrained optimal control problems applied to power split control for
  hybrid vehicles,'' \emph{Automatica}, vol.~50, no.~1, pp. 187--192, 2014.

\bibitem{butcher2016numerical}
J.~C. Butcher, \emph{Numerical methods for ordinary differential
  equations}.\hskip 1em plus 0.5em minus 0.4em\relax John Wiley \& Sons, 2016.

\bibitem{nazarmaster}
N.~Rozkvas, ``Real-time velocity and driving mode optimization for hybrid
  trucks using previewed road and traffic data,'' Master's thesis, Eindhoven
  University of Technology, 2019.

\bibitem{euvecto}
{European Commission Climate Action}, ``{V}ehicle {E}nergy {C}onsumption
  {c}alculation {TO}ol -- {VECTO},''
  \url{https://ec.europa.eu/clima/eu-action/transport-emissions/road-transport-reducing-co2-emissions-vehicles/vehicle-energy\_en},
  {A}ccessed: 08-11-2021.

\end{thebibliography}

\end{document}